\documentclass[aps,prl,twocolumn,superscriptaddress,showpacs,amsmath,amssymb]{revtex4}
\usepackage{epsfig}
\usepackage{bm}

\usepackage[T1]{fontenc}

\usepackage[usenames]{color}

\begin{document}
\title{Quantum heat fluctuations of single particle sources}

\author{F. Battista}

\affiliation{Division of Mathematical Physics, Lund University,
Box 118, S-221 00 Lund, Sweden}

\author{M. Moskalets}
\affiliation{Department of Metal and Semiconductor Physics, NTU ``Kharkiv Polytechnic Institute'', 61002 Kharkiv, Ukraine}

\author{M. Albert} 
\affiliation{Laboratoire de Physique des Solides, Universit\'e Paris Sud, 91405 Orsay, France.}

\author{P. Samuelsson}
\affiliation{Division of Mathematical Physics, Lund University,
Box 118, S-221 00 Lund, Sweden}

\begin{abstract} 
  Optimal single electron sources emit regular streams of particles,
  displaying no low frequency charge current noise. Due to the
  wavepacket nature of the emitted particles, the energy is however
  fluctuating, giving rise to heat current noise. We investigate
  theoretically this quantum source of heat noise for an emitter coupled 
  to an electronic probe in the hot-electron regime. The distribution
  of temperature and potential fluctuations induced in the probe is
  shown to provide direct information on the single particle
  wavefunction properties and display strong non-classical features.
\end{abstract}

\maketitle 

Recent years have witnessed a surge of interest in on-demand sources
for single particles in mesoscopic and nanoscale systems. This
interest was motivated by the experimental progress
\cite{highfreq,Feve,singlepar,Pekola,HBT,Hohls,giblin} on fast,
accurate single particle emitters, with operation frequencies reaching
the GHz regime. Fast and accurate emitters are key elements in the
efforts to obtain a quantum standard for the Ampere \cite{pekrev}. In
addition to metrological applications, coherent on-demand sources,
emitting regular streams of single particle wavepackets, are of great,
fundamental importance. As recently demonstrated \cite{HBT}, sources
implemented with edge states in the quantum Hall regime
\cite{Feve,Hohls} open up for quantum coherent few-electron experiments
\cite{Theorypump1, Theorypump2, Theorypump3,
  Theorypump4,Haack1,Haack2} as well as put in prospect quantum
information processing \cite{Janine,Chirolli,sherku} with clocked
single and entangled two-particle sources. Large efforts have also
been put into characterising the properties of on-demand sources via
the electrical current and its fluctuations
\cite{singem1,slava1,klevwf,haug,mahe,Albert,ours,mammut,Martin}.

Although the low-frequency charge emission of ideal on-demand sources
is noiseless, the emitted heat fluctuates \cite{heatmisha}. These
fluctuations are ubiquitous for quantum coherent sources; particles
emitted during a time shorter than the drive period ${\mathcal T}$
have an uncertainty in energy larger than $\hbar /{\mathcal
  T}$. Acting as emitters of quantum heat fluctuations, coherent
on-demand sources comprise ideal components for tests of heat
fluctuation relations \cite{Altland,Averin,Kung,Saira} in the quantum
regime \cite{Esposito} or for investigating the statistics of
temperature \cite{Heikilla,Laakso} or heat transfer \cite{Sanchez}
fluctuations in mesoscopic systems. In addition, the large versatility
of system parameters and pulse protocols \cite{HBT,Hohls} for
on-demand sources allows for a tailoring of the spectral properties of
the emitted wavepackets.

\begin{figure}[h]
\centerline{\psfig{figure=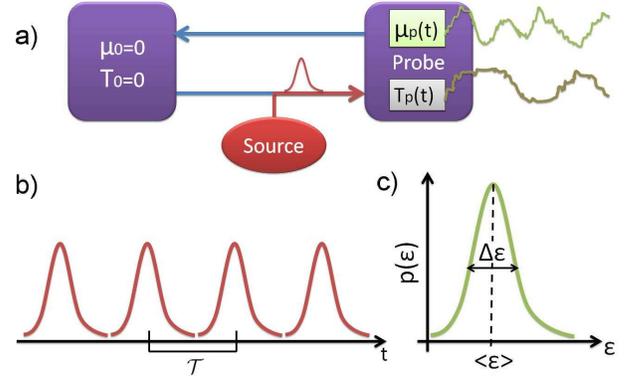,width=8cm}}
\caption{a) Schematic of an on-demand source injecting single particle wavepackets into an 
electronic probe, via the lower edge state of a conductor in the quantum Hall regime. 
 The probe is in the hot-electron regime, 
with a floating electron temperature $T_p(t)$ and chemical potential $\mu_p(t)$. 
Particles emitted from the probe flow along the upper edge into an electronic reservoir electrically 
grounded and kept at zero temperature.
b) Noiseless train of wavepackets, emitted from the single particle source with a frequency $\omega=2\pi/\mathcal{T}$. 
c) Probability distribution of energy $p(\epsilon)$ of the wavepacket, with average $\langle \epsilon\rangle$ and width 
$\Delta \epsilon$ shown.}
\label{fig1}
\end{figure}

In this work we provide a compelling illustration of the fluctuation
properties of system consisting of a generic coherent
on-demand source coupled to a hot-electron probe, see
Fig. \ref{fig1}. It is shown that the temperature and potential
fluctuations induced at the probe, besides fundamental constants,
depend only on the source frequency and the spectral properties of the
emitted wavepackets. For a wide range of parameters, the quantum
fluctuations are found to dominate over the classical ones. In
addition, the full distribution of the fluctuations reveals a direct
proportionality between the cumulants of the marginal temperature and
potential distributions, allowing for an experimental investigation of
the temperature fluctuation via correlators of the potential
fluctuations.

We first discuss the energy emission properties of an isolated,
optimal on-demand source, emitting a train of single particle
wavepackets $|\Psi\rangle$, equally spaced ${\mathcal T}=2\pi/\omega$
in time, see Fig. \ref{fig1}. The wavepackets, emitted on top of a
filled Fermi sea, are superpositions of states at different energies,
\begin{equation}\label{sourcestate}
|\Psi\rangle=\int_0^{\infty}d\epsilon c(\epsilon)\hat{b}^{\dagger}(\epsilon)|0\rangle,
\end{equation}
where $\hat{b}^{\dagger}(\epsilon)$ creates a particle at energy
$\epsilon>0$, $|0\rangle$ denotes the filled Fermi sea and
$c(\epsilon)$ an amplitude normalised as $\int d\epsilon
p(\epsilon)=1$, with $p(\epsilon)=|c(\epsilon)|^2$. 

Although energy emission for individual electrons can be accessed via
charge counting in weakly tunnel coupled systems
\cite{Averin,Saira,Sanchez}, for on-demand sources operating in the
GHz regime single-shot energy detection is presently not possible.
Instead one has to consider schemes where the effect of collecting a
large number $N\gg 1$ of emitted electrons, with a fluctuating total
energy $E$, become measurable. For an on-demand source characterized
by $p(\epsilon)$, the statistical distribution $P(E)$ of the total
energy can conveniently be written
\begin{eqnarray}\label{2}
 P(E)=\int d\xi e^{i\xi E+NF(\xi)}, \hspace{0.3cm} F(\xi)=\ln\left[\int d\epsilon p(\epsilon) e^{-i\xi \epsilon}\right]
\label{Estat}
\end{eqnarray}
where $NF(\xi)$ is the cumulant generating function for $P(E)$ and $N=t_0/{\mathcal T}$ is the
number of particles with $t_0 \gg {\mathcal T}$
the measurement time. The different cumulants of $P(E)$ are obtained
by successive derivatives of $ F(\xi)$ with respect to $\xi$, giving
for the average and the width
\begin{eqnarray}
\langle E \rangle=N\langle\epsilon\rangle, \hspace{0.3cm}(\Delta E)^2=N(\langle\epsilon^2\rangle-\langle\epsilon\rangle^2)\equiv N (\Delta \epsilon)^2
\label{Erels}
\end{eqnarray}
where $\langle .. \rangle=\int d\epsilon
.. p(\epsilon)$. Importantly, the direct relation 
between the statistics of $E$ and $p(\epsilon)$ of the individual wavepackets depends crucially
on the ideal operation of the source. Irregular wavepacket emission or
scattering in space or energy between emission and detection will make
$P(E)$ dependent on other factors. We stress that on-demand sources
which emit subsequent, identical electron and hole wavepackets
\cite{Feve}, (and hence no net charge) have the same heat emission
properties as pure electron sources with twice the drive frequency
$\omega$.

To access the heat fluctuation properties of the source we consider an
on-demand source coupled to a probe in the hot-electron regime, an
electrically and thermally floating terminal, see Fig. \ref{fig1}. The
source-probe setup is implemented in a conductor in the quantum Hall
regime, allowing to minimise scattering, elastically or inelastically,
between particle emission and collection in the probe. Particles
emitted from the source propagate to the probe along the lower edge
state. From the probe, emitted particles follow the upper edge and are
collected in a grounded electronic reservoir kept at low (here zero)
temperature.

In the hot-electron probe, injected particles thermalize rapidly, via
electron-electron interactions, on the time scale $\tau_{e-e}$. This
time is much shorter than the typical time $\tau_d$ a particle spends
inside the probe before being reemitted. However, the energy exchange
with the lattice phonons takes place on the time scale $\tau_{e-ph}$,
much longer than $\tau_d$. The electron distribution in the probe is
then in a quasi-equilibrium state, characterized by a chemical
potential and temperature. In order to prevent charge and energy
pile-up in the floating probe, both the chemical potential $\mu_p(t)$
and temperature $T_p(t)$ develop fluctuations in time. The potential
fluctuations can be detected by present day electrical
measurements. Importantly, as we now show, the potential fluctuations
also provide direct information on the quantum heat fluctuation of the
source, via $p(\epsilon)$ of the individual particles.

To present a clear and compelling picture we analyse the temperature
and potential fluctuations within a Boltzmann-Langevin approach. We
write $T_p(t)=\bar{T}_p+\delta T_p(t)$ and
$\mu_p(t)=\bar{\mu}_p+\delta \mu_p(t)$ with $\bar{T}_p,\bar{\mu}_p$
average quantities and $\delta T_p(t),\delta \mu_p(t)$ fluctuating
Langevin terms.  The statistics of $\delta T_p(t)$ and $\delta
\mu_p(t)$ is determined from underlying quantum properties, as
discussed below. First we consider the average quantities $\bar{T}_p$
and $\bar{\mu}_p$.  The starting point is the operator for charge
current at the probe \cite{Butt}, $\hat{I}_p^c(t)=\int d\epsilon
d\epsilon'
e^{i(\epsilon-\epsilon')t/\hbar}\hat{i}_p(\epsilon,\epsilon')$ where
$\hat{i}_p(\epsilon,\epsilon')=(e/h)[\hat{b}^{\dagger}(\epsilon)\hat{b}(\epsilon')-\hat{a}^{\dagger}(\epsilon)\hat{a}(\epsilon')]$
with $\hat{b}^{\dagger}(\epsilon) [\hat{a}^{\dagger}(\epsilon)]$
creating particles incident on [emitted from] the probe. By analogy we
write the operator for the heat current $\hat{I}_p^h(t)=\int d\epsilon
d\epsilon'
e^{i(\epsilon-\epsilon')t/\hbar}[(\epsilon+\epsilon')/2-\mu_0]\hat{i}_p(\epsilon,\epsilon')$,
where $\mu_0=0$ is the chemical potential of the reservoir.  Taking
the quantum average with respect to the emitted source state,
Eq. (\ref{sourcestate}), and the state of the probe, and averaging
over a time much longer than the period $\mathcal{T}$, we arrive at
the dc component of charge and heat currents
\begin{equation}\label{curreq}
 \langle I_p^c\rangle=\sigma \frac{e}{\mathcal T}-g_0\frac{\bar \mu_p}{e}, \hspace{0.3cm} \langle I_p^h\rangle=\frac{\langle \epsilon \rangle}{\mathcal T}-\frac{g_0}{2}\left[\frac{\bar \mu_p^2}{e^2}+l_0\bar T_p^2\right] 
\end{equation}
where $g_0=e^2/h$ is the (single spin) conductance quantum and
$l_0=(\pi k_B/e)^2/3$ the Lorentz number. To account for both types of
sources discussed, we introduced $\sigma=0$ for sources emitting no
net charge and $\sigma=1$ for sources emitting one electron
per cycle.  We note that the first and the second terms of $\langle
I_p^c\rangle$ and $\langle I_p^h\rangle$ in Eq. (\ref{curreq}) are the
currents emitted by the source and the probe respectively.  The
conditions for zero average charge and heat currents at the probe, $\langle
I_p^c\rangle=0$ and $\langle I_p^h\rangle=0$, give from
Eq. (\ref{curreq})
\begin{equation}\label{av}
\bar{\mu}_p=\sigma\hbar\omega, \hspace{0.3cm} \bar{T}_p=\sqrt{\frac{1}{g_0l_0\mathcal{T}}[2\langle \epsilon\rangle-\sigma\hbar\omega]}.
\end{equation}
Importantly, $\bar{\mu}_p$ and $\bar{T}_p$ depend only on the source
properties $\omega$ and $\langle \epsilon \rangle$ and fundamental
constants \cite{HBT}. We note that $\langle
\epsilon\rangle>\hbar\omega/2$ follows along the lines of
Ref. \cite{avron}.

Turning to the temperature and chemical potential fluctuations
\cite{Butrev}, the quantities of primary experimental interest are the
low frequency correlators $\langle (\delta
\mu_p)^2\rangle\equiv(1/t_0)\int dtdt'\langle \delta \mu_p(t) \delta
\mu_p(t')\rangle$ and equivalently for $\langle (\delta
T_p)^2\rangle$, with the measurement time $t_0\gg\mathcal{T}$.  We
first point out that the total fluctuations of charge and heat
currents $\Delta I_p^c$ and $\Delta I_p^h$ are made up by bare
fluctuations $\delta I_p^c$ and $\delta I_p^h$ and fluctuations due to
the varying temperature and voltage of the probe,
$\partial_{T_p}\langle I_p^{x}\rangle\delta T_p$ and
$\partial_{\mu_p}\langle I_p^{x}\rangle\delta \mu_p$, with $x=c,h$, as
\begin{equation}\label{flucsys}
  \left(\begin{array}{c}\Delta I_p^c \\ \Delta I_p^h \end{array}\right)=\left(\begin{array}{c}\delta I_p^c \\ \delta I_p^h \end{array}\right)+\left(\begin{array}{cc} \partial_{\mu_p}\langle I_p^{c}\rangle & \partial_{T_p}\langle I_p^{c}\rangle \\ \partial_{\mu_p}\langle I_p^{h}\rangle & \partial_{T_p}\langle I_p^{h}\rangle\end{array}\right)\left(\begin{array}{c} \delta \mu_p \\ \delta T_p \end{array}\right)
\end{equation}
suppressing for shortness the time dependence of the
fluctuations. From Eqs. (\ref{curreq}) and (\ref{flucsys}), taking
into account the conservation of charge and heat current fluctuations,
{\it i.e.}  $\Delta I_p^c=0$ and $\Delta I_p^h=0$, we can express
$\delta \mu_p$ and $\delta T_p$ in terms of the bare charge and heat
fluctuations as
\begin{equation}
  \delta \mu_p=\frac{h}{e} \delta I^c_p, \hspace{0.3cm} \delta T_p=\frac{1}{g_0l_0\bar{T}_p}\left(\delta I^h_p-\frac{\bar{\mu}_p}{e}\delta I^c_p\right).
\end{equation}
The correlators $\langle (\delta \mu_p)^2\rangle$ and $\langle (\delta
T_p)^2\rangle$ can be thus expressed in terms of low frequency
correlators of bare charge and heat fluctuations $\langle\delta
I_p^x\delta I_p^y\rangle\equiv(1/t_0)\int dtdt'\langle\delta
I_p^x(t)\delta I_p^y(t')\rangle$. The correlator of the Langevin
terms $\langle\delta I_p^x(t)\delta I_p^y(t')\rangle$ is evaluated by
taking the quantum average of the corresponding correlator of current
operators $\hat{I}_p^c$, $\hat{I}_p^h$ following Ref. \cite{Butt}. We
arrive at
\begin{eqnarray}\label{fluc}
\langle(\delta \mu_p)^2\rangle&=&hk_b\bar{T}_p,\\ \nonumber
\langle (\delta T_p)^2\rangle&=&\frac{1}{g_0l_0}\left[k_b\bar{T}_p+\frac{1}{2}\frac{(\Delta\epsilon)^2}{\langle \epsilon\rangle-\sigma \hbar \omega/2}\right].\label{corrdeltav3}
\end{eqnarray}
The potential fluctuations $\langle(\delta \mu_p)^2\rangle$ are
proportional to the average temperature, typical for equilibrium
systems \cite{Butrev}. In contrast, the temperature fluctuations
$\langle(\delta T_p)^2\rangle$ are a sum of two physically distinct
terms. The first term, the classical fluctuations, is proportional to
$\bar{T}_p$ and results from the finite temperature of the probe and
would be present even if the injected particles had a well defined
energy, \textit{i.e.} $\Delta \epsilon=0$.  The second term, quantum
fluctuations, is proportional to $(\Delta \epsilon)^2$ and is a direct
result of the uncertainty of the energy of the injected
particle. Importantly, for a broad range of drive frequencies $\omega$
and wavepacket mean energies $\langle \epsilon \rangle$ and widths
$\Delta \epsilon$, the quantum term dominates over the classical
one. We also point out that there are no correlation between the
voltage and the temperature fluctuations, \textit{i.e.}
$\langle\delta\mu_p\delta T_p\rangle=0$.

In order to investigate the presence of quantum heat fluctuations in
higher order potential correlations, and also to provide a complete
picture of the temperature and potential fluctuations, we turn to the
full probability distribution.  To relate to the average and
fluctuation correlators above, we introduce a dimensionless potential
$\mu=(1/h)\int_0^{t_0}dt\mu_p(t)$, and temperature
$T=(1/h)\int_0^{t_0}dt k_bT_p(t)$, fluctuating quantities integrated
over the measurement time $t_0$.  The joint probability distribution
$\mathcal{P}_{t_0}(\mu, T)$ can be conveniently written in terms of a
cumulant generating function $G(\chi,\theta)$ as
\begin{equation}\label{ptog2}
 \mathcal{P}_{t_0}(\mu,T)=\frac{1}{(2\pi)^{2}}\int d\chi\int d\theta e^{-i\theta T-i\chi\mu+G(\chi,\theta)},
\end{equation}
with $\chi$ and $\theta$ counting fields for $\mu$ and
$T$ respectively. 
From  $G(\chi,\theta)$ the low frequency cumulants are then, by construction, obtained from
successive derivatives with respect to the counting fields, giving
$t_0\langle \delta T_p^n\delta\mu_p^m\rangle=(-ih)^{n+m}k_b^{-n}\partial^n_{\theta}\partial^m_{\chi}G(\chi,\theta)|_{\chi,\theta=0}$.

To determine $G(\chi,\theta)$ we first spell out the relations between
the time scales in the problem. The potential $\mu_p(t)$ fluctuates on
the time scale given by the RC-time, $\tau_{RC}$, while the
temperature $T_p(t)$ typically fluctuates on the time scale of the
dwell-time in the probe, $\tau_{d}$.  We assume that the system is in
the limit $\tau_{e-e}\ll\tau_{RC},\tau_{d}\ll\tau_{e-ph}$.  Moreover,
we consider periods of the source, $\mathcal{T}$, and measurements
time such that $t_0\gg\tau_{d},\tau_{RC}\gg\mathcal{T}$. On time
scales $\tau$ such that $\mathcal{T}\ll\tau\ll\tau_{d},\tau_{RC}$ the
statistics of net transferred energy $E_p$ and charge $Q_p$ in the
probe can be described by the source generating function $\tau
h_s(\lambda,\xi)$ with
\begin{equation}
h_s=\frac{\omega}{2\pi}\left[-ie\sigma\lambda+F(\xi)\right]
\end{equation}
with $F(\xi)$ given in Eq. (\ref{Estat}) and the probe generating function $\tau h_p(\lambda,\xi,E_p,Q_p)$ with \cite{lev, kinpil, pilgram}
\begin{eqnarray}
h_p=\frac{1}{h}\int d\epsilon \bigg[\ln[1+f_p(\epsilon)(e^{ie\lambda+i\epsilon\xi}-1)]+\\
\ln[1+f(\epsilon)(e^{-ie\lambda-i\epsilon\xi}-1)]\bigg],\nonumber
\end{eqnarray}
where $f_p(\epsilon)=f_p(\epsilon,\mu_p,T_p)$ and $f(\epsilon)$ are
the probe and the reservoir distribution functions (see Fig. \ref{fig1}) and $\xi$ and
$\lambda$ are the counting fields for $E_p$ and $Q_p$
respectively. The energy $E_p$ and charge $Q_p$ are related to $T_p$
and $\mu_p$ as $E_p= \nu[\mu_p^2/2+ (\pi k_b T_p)^2/6 ]$ and $Q_p=\nu
e \mu_p$, where $\nu$ is the density of states in the probe.  Working
within the framework of the stochastic path integral formalism
\cite{pilgrambutt, pilgram}, we can then express $G(\chi,\theta)$ as a
path integral over all configurations of $E_p$ and charge $Q_p$ during
the measurement. In the long time limit we have
\begin{equation}\label{w}
e^{G(\chi,\theta)}=\int dQ_p dE_p d\lambda d\xi e^{S(Q_p,E_p,\lambda,\xi)}
\end{equation}
where $S(Q_p,E_p,\lambda,\xi)=t_0[i \theta
k_bT_p/h+i\chi\mu_p/h+h_p(Q_p,E_p,\lambda,\xi)+h_s(\lambda,\xi)]$.
Similar to Refs. \cite{kinpil, pilgram,Heikilla}, the integral in
Eq. (\ref{w}) is solved in the saddle point approximation. Inserting
the solutions for $Q_p,E_p,\lambda,\xi$ into $S(Q_p,E_p,\lambda,\xi)$
we arrive at
\begin{eqnarray}\label{genfun}
  G(\chi,\theta)=N\bigg[\frac{d[zF(z)]}{dz}+\sigma(z+i\chi)\bigg]
\label{g0exact}
\end{eqnarray}
recalling that $N=t_0/\mathcal{T}$, where $F(z)\equiv
F(\xi)|_{i\xi=z/\hbar\omega}$ and $z$ is found from the relation
$z^2[dF/dz+\sigma/2]=-(\pi^2/6)(1-\sqrt{1-2g})^{2}$ with
$g(\chi,\theta)=(3/\pi^2)[(i\chi)^2/2+i\theta]$. From
Eq. (\ref{g0exact}) we note several important things. First, by
expanding $G(\chi,\theta)$ in terms of $\chi$ and $\theta$ we see that
the first two cumulants reproduce the results in Eqs. (\ref{av}) and
(\ref{fluc}).  Second, all the even chemical potential cumulants, from
the first two terms in Eq. (\ref{g0exact}), can be expressed in terms
of the temperature cumulants as
 \begin{equation}
  \langle(\delta \mu_p)^{2n}\rangle=(2n-1)!!(k_bh)^n\langle(\delta T_p)^n\rangle                                              
\end{equation}
a consequence \cite{Flindt} of the counting fields entering via $g(\chi,
\theta)$. 

The full distribution $\mathcal{P}_{t_0}(\mu,T)$ can be found (to
exponential accuracy) by solving the integral in Eq. (\ref{ptog2}) in
the saddle point approximation. We obtain the compact expression
\begin{equation}\label{pta}
 \ln \mathcal{P}_{t_0}(\mu,T)=-iT\theta^*+G(0,\theta^*)-(\mu-\bar{\mu})^2/(2T)
\end{equation}
where $\bar{\mu}=t_0\bar{\mu}_p/h=\sigma N$ and the saddle point
solution $\theta^*$ is found from the relation
$dF/dz|_{z=z^*}+\sigma/2=-(\pi^2/6)(Tq^{*}/N)^2$, with
$z^*=N(q^{*}-1)/(q^{*}T)$ and $q^{*}=\sqrt{1-i6\theta^*/\pi^2}$.
Importantly, Eq. (\ref{pta}) shows that the potential $\mu$ displays
Gaussian fluctuations, of width $\sqrt{T}$, around the average
$\bar{\mu}$ for any given temperature $T$. Hence, the marginal potential distribution $\mathcal{P}_{t_0}(\mu)=\int dT \mathcal{P}_{t_0}(\mu,T)$ is symmetric around $\bar \mu$, albeit not Gaussian. Moreover, the marginal 
distribution for the temperature $\mathcal{P}_{t_0}(T)=\int d\mu
\mathcal{P}_{t_0}(\mu,T)$ is given by $\ln
\mathcal{P}_{t_0}(T)=-iT\theta^*+G(0,\theta^*)$.

\begin{figure}[h]
\centerline{\psfig{figure=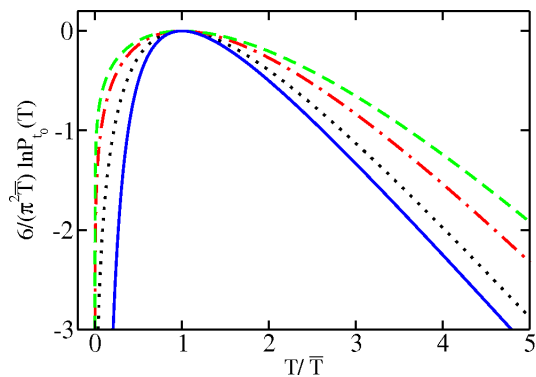,width=7cm}}
\caption{Normalised logarithm of the probability distribution $\mathcal{P}_{t_0}(T)$ as a function of $T/\bar{T}$, with $\bar{T}$ the average value of
$T$. The curves correspond to the narrow Gaussian wavepacket energy distribution (blue solid) in Eq. (\ref{Pgauss}) and to an exponential
energy distribution with $\alpha=1$ (black dotted), $3$ (red dash-dotted), $5$ (green dashed). See text for details.}
\label{plot}
\end{figure}

To illustrate our results we evaluate $\ln \mathcal{P}_{t_0}(T)$ for
two distinct cases. First, as a reference, we consider the generic
Gaussian spectral distribution
$p(\epsilon)=1/(\sqrt{2\pi}\Delta\epsilon)
e^{-(\epsilon-\langle\epsilon\rangle)^2/(2(\Delta\epsilon)^2)}$. Taking
the classical limit, with $\Delta \epsilon \ll \langle \epsilon
\rangle$, we get the simple result
\begin{equation}
 \ln \mathcal{P}_{t_0}(T)=-(\pi^2/6)T\left(1-\bar{T}/T\right)^2
\label{Pgauss}
\end{equation}
with $\bar{T}=t_0k_b\bar{T}_p/h$ the average value of $T$. The log
probability is plotted in Fig. \ref{plot}. For small fluctuations
$T-\bar T \ll \bar T$ the distribution is Gaussian while for $T \ll
\bar T$ the probability is suppressed $\mathcal{P}_{t_0}(T)\propto
e^{-\pi^2\bar T^2/(6T)}$, guaranteeing
$\mathcal{P}_{t_0}(T)\rightarrow 0$ for $T \rightarrow 0$. The
probability for large fluctuations $T\gg \bar T$ is suppressed as
$\mathcal{P}_{t_0}(T)\propto e^{-T\pi^2/6}$. For finite but small
width $\Delta \epsilon \ll \langle \epsilon \rangle$, the log
probability in Eq. (\ref{Pgauss}) is multiplied by the term $1+(\Delta
\epsilon/\langle \epsilon \rangle)^2\bar T^2\pi^2/(12 TN)$ for $T \sim
\bar T$, a small quantum correction.
In contrast, for an exponential distribution
$p(\epsilon)=(1/\langle\epsilon\rangle) e^{-
  \epsilon/\langle\epsilon\rangle}$ derived in Ref. \cite{klevwf} for
adiabatic particle emission and investigated in Ref. \cite{singem1}, we find the probability
\begin{equation}
 \ln \mathcal{P}_{t_0}(T)=-\frac{\pi^2\bar T}{6}\left(\frac{T}{\bar T}(1-q^*)-\frac{2}{\alpha}\ln\left[\frac{Tq^*}{\bar T}\right]\right)
\label{PFeve}
\end{equation}
where $q^*=(\alpha/2+\sqrt{\alpha\bar T/T+1+\alpha^2/4})/(T/\bar
T+\alpha)$ and $\alpha=\pi\sqrt{\langle \epsilon
  \rangle/(6\hbar\omega)}$. The probability distribution is plotted in
Fig. \ref{plot} for different values of $\alpha$. We see that for
increasing $\alpha$, corresponding to slower drive $\omega$ and/or
larger average wavepacket energies $\langle \epsilon \rangle$, the
distribution gets increasingly broad and deviates strongly from the
classical one in Eq. (\ref{Pgauss}).

In conclusion we have investigated the quantum fluctuations of the
heat current emitted from a single particle source. We show that these
quantum heat fluctuations can be detected via electrical potential
fluctuations of a probe coupled to the source. For typical parameters
\cite{Feve} $2\pi \omega \sim 1GHz$ and $\langle \epsilon \rangle \sim
0.1 meV$ we get $\bar \mu_p \sim 5 \mu eV$ and $\bar T_p=0.2K$,
demonstrating the experimental feasibility of our proposal.

We acknowledge valuable discussions with C. Flindt and
M. B\"uttiker. We acknowledge support from the Swedish VR.

\end{document}